\documentstyle[proceedings,namedreferences,epsf]{crckapb}

\newcommand{\md}{\mbox{d}}

\begin{opening}
\title{On the nonthermal X-ray emission in blazar jets}
\author{F.M. Rieger}
\institute{Universitaets-Sternwarte\\
Geismarlandstr. 11, 37083 Goettingen}
\author{J.G. Kirk}
\institute{Max-Planck-Institut f\"ur Kernphysik\\
 Postfach 10 39 80, 69029 Heidelberg}
\author{A. Mastichiadis}
\institute{University of Athens, Department of Physics\\
Panepistimiopolis, GR-15783 Zografos, Greece}
\end{opening}
\begin{document}
\section{Introduction}
   The origin of the nonthermal continuum emission in Blazars is a challenging problem
   in the area of active galactic nuclei. The X-ray emission is usually interpreted 
   as synchrotron radiation of relativistic electrons accelerated in a jet which itself moves 
   at relativistic speed towards the observer, while the TeV gamma-rays are though to arise 
   via synchrotron self-comptonisation (e.g. Mastichiadis \& Kirk~1997), via comptonisation 
   of external photons (e.g. Dermer, Schlickeiser, Mastichiadis~1992) or via the decay of 
   relativistic protons (e.g. Mannheim~1998).\\
   In this contribution, we present a self-consistent model, where electrons experience 
   first-order Fermi acceleration at a shock front in a relativistic jet and radiate 
   synchrotron emission in a post-shock region which contains a homogeneous magnetic 
   field. The full time, space and momentum dependence of the electron distribution 
   function is used for a calculation of the nonthermal synchrotron spectra. 
   By varying the rate at which electrons enter the acceleration process, we calculate 
   the time-dependence of the spectral index for the X-ray emission predicted in this 
   model. The results show that the synchrotron spectral index displays a characteristic 
   looplike behaviour with intensity (as has been observed in several blazars), where 
   the orientation of the loop depends on whether the acceleration time scale is 
   comparable to the synchrotron cooling time scale or not. We show that our model 
   provides a good fit to the observed evolution of the spectral index of 
   Mkn 421 during a flare in 1994.

\section{Acceleration model}
   We assume diffusive particle acceleration at a shock front moving at a constant 
   speed $u_s$ along a cylindrically symmetric jet. Following Ball \& Kirk (1992) we 
   divide the jet into an acceleration region close to the shock and a cooling region 
   downstream (see figure). Electrons in the acceleration region repeatedly encounter a plane 
   shock. We assume that high energy particles pass unaffected through the shock and scatter 
   elastically of magnetic irregularities (Alfv\'en waves) in the fluid on either side of the 
   shock. In the rest frame of the shock the speed of the upstream fluid exceeds the downstream 
   fluid speed so that particles gain energy by the first-order Fermi process (Drury 1983). 
   Electrons which escape from the acceleration region into the downstream region
   cease to diffuse and are assumed to remain frozen into the fluid flow,
   suffering only synchrotron losses.\\
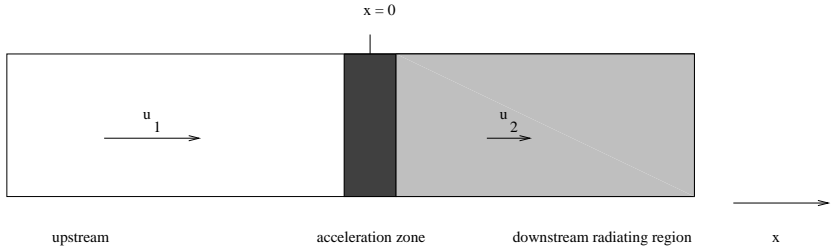
   \begin{figure}[htb]
       \vspace{-0.5cm}
       \begin{center}
          \epsfxsize11.0cm
          \mbox{ 
       \epsffile{aas.f1}}
       \end{center}
       \vspace{0cm}
   \end{figure}

   Consider electrons which are injected and enter the acceleration mechanism at the 
   constant rate $Q$ with Lorentz factor $\gamma_0$. Although it is possible to 
   write down the distribution function of such particles as a function of momentum,
   space and  time (e.g. Toptygin~1980, Drury~1991), we use a spatially averaged 
   model (Axford~1981, Bogdan \& V\"olk~1983), in which an electron is presumed
   to undergo continuous acceleration while in the vicinity of the shock, such that
   its Lorentz factor $\gamma$ increases at a rate $\gamma/t_{\rm acc}$, where 
   $t_{\rm acc}^{-1}=D/t_c$ is the acceleration rate and $D$ is the average 
   fractional momentum gain per cycle (for crossing and recrossing the shock), and 
   $t_c$ is the average time taken to perfom such a cycle (see Bell~1978, Drury~1983, 
   Kirk, Melrose, Priest~1994). The acceleration process is coupled with the fact 
   that particles in the downstream region have a chance of not returning to the shock. 
   These particles escape at a rate $t_{\rm esc}^{-1}$, which is 
   just the escape probability per cycle $P_{\rm esc}$ divided by the average 
   cycle time $t_c$. Assuming pitch angle scattering is sufficiently rapid to keep
   the distribution function almost isotropic, we can easily include the 
   electron synchrotron loss term $\langle \dot \gamma \rangle_s = -\beta_s\,
   \gamma^2$ in a differential equation governing the number $N(\gamma)\md\gamma$ 
   of particles in the acceleration zone around the shock with Lorentz factor 
   between $\gamma$ and $\gamma + \md\gamma$ (e.g. Kirk, Melrose, Priest~1994, 
   Mastichiadis \& Kirk~1997):
   \begin{equation}\label{phano}
   \frac{\partial N}{\partial t} + \frac{\partial}{\partial \gamma} \Bigl[\left(
   \frac{\gamma}{t_{\rm acc}} -\beta_s\,\gamma^2 \right) N \Bigr] +
   \frac{N}{t_{\rm esc}} = Q \delta (\gamma - \gamma_0)
   \end{equation}
   where $\beta_s= 4\,\sigma_{\rm T}\, B^2/(6\,\mu_0\,m_{\rm e}\,c)$
   with $\sigma_{\rm T}= 6.65 \cdot 10^{-29} {\rm m}^2$ the 
   Thomson cross section and $B$ the magnetic field in Tesla.
   There obviously exists an important Lorentz factor in this problem: at the
   point $\gamma_{\rm max}=1/(\beta_s\,t_{\rm acc})$ acceleration is exactly 
   balanced by synchrotron losses. It is to be expected that the solution always
   vanishes for $\gamma > \gamma_{\rm max}$ and also for Lorentz factors to which there 
   has not been enough time to accelerate particles.\\
   From the synchroton losses we get a typical cooling time scale given 
   by $t_{\rm cool}=1/(\gamma\,\beta_s)$. 
   As $\gamma \rightarrow \gamma_{\rm max}$, $t_{\rm cool}$ tends to $t_{\rm acc}$.\\ 
   In the model developed by Ball \& Kirk~(1992), accelerated particles escape 
   into the downstream plasma, where they radiate. 
   We can formulate the kinetic equation obeyed by the density of particles in 
   the radiation zone most compactly using a coordinate system at rest in the 
   radiating plasma. Note that in this case the radiation must be Doppler-boosted to
   the observer's frame: 
   quantities such as the intensity or the flux of the synchrotron radiation, which are 
   computed in the rest frame of the source (plasma rest system), require a Lorentz 
   transformation into the observer's frame, determined by the Doppler factor $\delta= 
   1/[\Gamma (1-\beta\,\cos \theta)]$, where $\Gamma=(1-u_2^2/c^2)^{-1/2}$, 
   $\beta=u_2/c$ and $\theta$ is the angle between the direction of motion and the line 
   of sight to the observer.\\
   In the plasma rest system the shock front provides a moving source of electrons, 
   which subsequently suffer energy losses, but are assumed not to be transported 
   in space. The kinetic equation governing the differential density 
   $\md n(x,\gamma,t)$ of particles in the range $\md x$, $\md \gamma$ then is 
   \begin{eqnarray}\label{trans}
   \frac{\partial n}{\partial t} -
   \frac{\partial}{\partial \gamma}(\beta_{\rm s}\,\gamma^2\,n) &=&
   \frac{N(\gamma,t)}{t_{\rm esc}}\delta(x-x_{\rm s}(t)) 
   \end{eqnarray} where $x_{\rm s}(t)$ is the position of the shock front at time 
   $t$. Equations (\ref{phano}) and (\ref{trans}) have been solved analytically (see
   Kirk, Rieger, Mastichiadis~1998). To obtain the synchrotron emissivity as a function 
   of position, time and frequency, it remains to convolve the density $n$ with the
   synchrotron Green's function $P(\nu,\gamma)$ (e.g. Melrose~1980). 
   Then at a point  $x=X$ ($>u_{\rm s}t$) on the symmetry axis of the source at 
   time $t$, the specific intensity of radiation in the $\vec{x}$ direction depends 
   on the retarded time ${\bar t}=t-X/c$ and is given by
   \begin{equation}
   I(\nu,{\bar t})\,=\,\int\md\gamma P(\nu,\gamma)\int\md x
   \,n(x,\gamma,{\bar t}+x/c)
   \end{equation}

\section{Spectral variability}
\begin{figure}[htb]
       \vspace{-0cm}
       \begin{center}
          \epsfxsize9.0cm 
          \mbox{ \epsffile{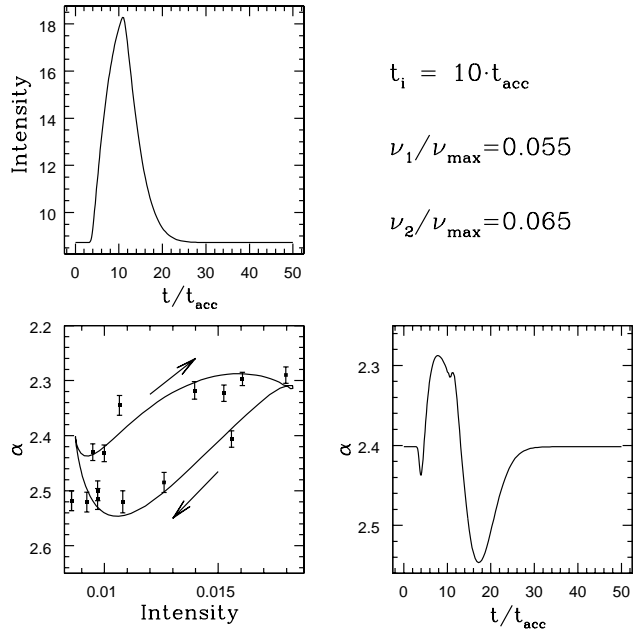}}
       \end{center}
       \vspace{-0cm} 
       \caption{Evolution of the photon index $\alpha$ and intensity (arbitrary units) 
          in a flare at low frequencies. The loop in the photon index $\alpha$ is 
          followed in the clockwise direction. The Mkn 421 data points in this picture 
          are taken from Takahashi et al.~1996 (see text for details).}
       \label{loop1}
   \end{figure}

\begin{figure}[htb]
       \vspace{-0cm}
       \begin{center}
          \epsfxsize9.0cm 
          \mbox{ \epsffile{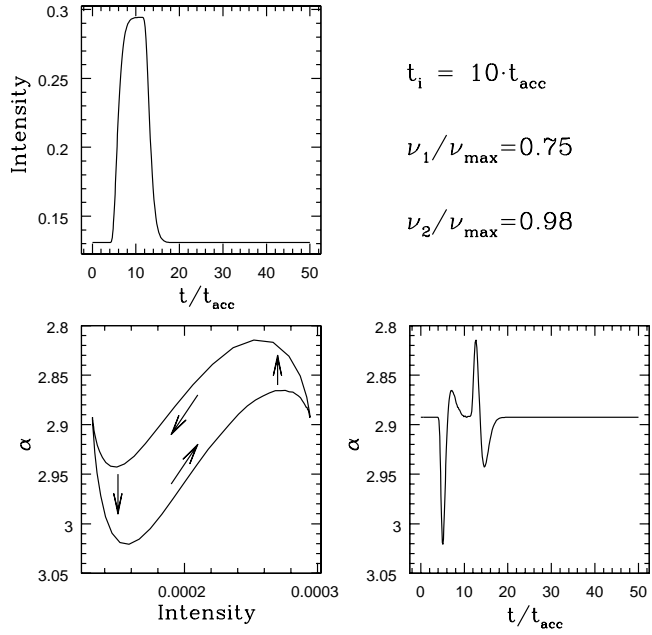}}
       \end{center}
       \vspace{-0.5cm} 
       \caption{Evolution of the photon index $\alpha$ and intensity (arbitrary units) in 
        the same flare as plotted in Fig.(\ref{loop1}) if the system is observed near the 
        maximum frequency $\nu_{\rm max}$. The loop in the photon index $\alpha$ is now 
        followed in the anti-clockwise direction.}
       \label{loop2}
   \end{figure}

   \vspace{-0.2cm}
   The model described above provides a remarkably good fit to the radio to X-ray spectra of 
   several blazars. In a recent publication (Kirk, Rieger, Mastichiadis~1998) we apply 
   our model to the nonthermal X-ray emission of Mkn 501, where we get a very good 
   representation of the observed synchrotron spectrum. Since we solve Eq. (\ref{phano}) 
   and Eq. (\ref{trans}) for the full time, space and momentum dependence of the electron 
   distribution function, we are able to apply our model to the observed variability or 
   flaring behaviour in blazars. Variability or flaring can arise for several reasons, 
   e.g. when the shock front overruns a region where the local plasma density is 
   enhanced. In this case the number of particles entering the acceleration process 
   might increase for a time $t_i$. A simple representation of a flare, for example, 
   is obtained by setting $Q(t)=Q_0$ for $t<0$ and $t>t_i$ and 
   $Q(t)=Q_1$ for $0<t<t_i$.
   In Fig. (\ref{loop1}) we plot the corresponding synchrotron emission given in the 
   rest frame of the downstream plasma for the case $Q_1=1.8\,Q_0$. Well below the maximum 
   emitting frequency, where the acceleration time scale $t_{\rm acc}$ is much shorter than the 
   synchrotron cooling time scale $t_{\rm cool}$, the spectrum shows the typical soft 
   lag behaviour observed in several sources (e.g. PKS 2155-304: Sembay 
   et al.~1993, H0323-022: Kohmura et al.~1994, Mkn 421: Takahashi et al.~1996): the 
   spectrum becomes softer in declining phases and harder during phases of rising flux, 
   which indicates that variations in the hard X-rays always lead those in the soft
   X-rays. An interpretation of this behaviour has been given in terms of synchrotron 
   cooling (e.g. Tashiro et al.~1995). In this case, information about changes in the
   injection propagates from high energy, where the cooling process is faster, to low 
   energies.\\ 
   In Fig. (\ref{loop1}) we also give a qualitative fit for the evolution of the 
   photon index $\alpha$ for Mkn 421 during an X-ray flare, observed with the X-ray 
   satellite ASCA in May 1994 (note that the data should be treated with caution, since
   they were derived for the physically unlikely case that the absorbing column $N_H$ was 
   allowed to vary, see Takahashi et al.~1996). For the plot we use both the constraints 
   given in Takahashi et al. (1996) and the parameters obtained by Mastichiadis \& Kirk 
   (1997) in fitting the observed multiwavelength spectrum of Mkn 421 in 1994. 
   The observed maximum frequency is roughly $\nu_{\rm max} \approx 1.4\cdot 10^{18}$Hz,
   so that, applying a Doppler factor of $15$, the maximum electron Lorentz factor is $10^6$. 
   The results are plotted in Fig. (\ref{loop1}) for a shock velocity $u_s=0.26\,c$ and for
   frequencies of observation well below the the maximum frequency.    
   Closer to the maximum frequency, where the acceleration and cooling time scale   
   become comparable, the picture changes. In Fig. (\ref{loop2})
   we show the predicted results for the same parameters as used in Fig. (\ref{loop1}),
   for a frequency of observation close to the maximum frequency. Here the loop in the 
   photon index $\alpha$ is traced in the anti-clockwise direction. This behaviour is 
   a typical signature of the underlying acceleration mechanism. As we have argued 
   recently (Kirk, Rieger, Mastichiadis~1998), this feature represents the effect that now 
   information about the occurrence of a flare propagates from low to high energy as 
   particles are gradually accelerated.\\
   Although in our model we neglect factors such as, for example, inhomogenities in the 
   source geometry, the result that the loop is traced in a clockwise direction if 
   controlled by synchrotron cooling and followed in the opposite, i.e. anti-clockwise
   direction if dominated by the acceleration time scale is robust.\\ \\ \\
   {\small{Acknowledgements: F.M.R. acknowledge support under DFG Ma 1545/2-1. The 
   collaboration between the MPK and the University of Athens is supported by the 
   European Commission under the TMR Programme, contract FMRX-CT98-0168.}}

   \newpage

\vspace{-0.5cm}
\section{References}
\begin{quote}
{\small
Axford W.I., 1981, {\it Proc. 17th. Internat. Cosmic Ray Conf.} (Paris), 12, 155\\
Ball L., Kirk J.G. 1992, {\it ApJ} 396, L39\\
Bell A.R. 1978, {\it MNRAS} 182, 147\\
Bogdan T.J., V\"olk H.J. 1983, {\it A\&A} 122, 129\\
Dermer C.D., Schlickeiser R., Mastichiadis A. 1992, {\it A\&A} 256, L27\\
Drury, L.O'C. 1983, {\it Rep. Progr. Phys.} 46, 973\\
Drury, L.O'C., 1991, {\it MNRAS} 251, 340\\
Catanese M., et al. 1997, {\it ApJ Letters} 487, L143\\
Kirk J.G., Melrose D.B., Priest E.R. 1994, {\it Plasma Astrophysics},
eds. A.O. Benz, T.J.-L. Couvoisier, Springer, Berlin\\
Kirk J.G., Rieger F.M., Mastichiadis A. 1998, {\it A\&A} 333, 452\\
Kohmura Y., et al. 1994, {\it PASJ} 46, 131\\
Mannheim K. 1998, {\it Science} 279, 684\\
Mastichiadis A., Kirk J.G., 1997, {\it A\&A}, 320, 19\\
Melrose D.B. 1980, {\it Plasma Astrophysics}, Vol. I, Gordon \& Breach, New York\\
Sembay S., et al. 1993, {\it ApJ} 404, 112\\
Takahashi T., et al. 1996, {\it ApJ} 470, L89\\
Tashiro M., et al. 1995, {\it PASJ} 47, 131\\
Toptygin I.N. 1980, {\it Space Sci. Rev.} 26, 157\\

}
\end{quote}

\end{document}